\documentclass[11pt,reqno]{article}
\usepackage{latexsym}
\usepackage{url,citesort}
\usepackage[thmmarks]{ntheorem}

\usepackage{maple2e}
\usepackage{amsmath,amssymb,amsfonts,amscd,graphicx}

\renewcommand{\theequation}{\thesection.\arabic{equation}}

\let\ssection=\section

\renewcommand{\section}{\setcounter{equation}{0}\ssection}

\newcounter{mnotecount}[section]

\renewcommand{\themnotecount}{\thesection.\arabic{mnotecount}}

\newcommand{\mnote}[1]
{\protect{\stepcounter{mnotecount}}$^{\mbox{\footnotesize $
\bullet$\themnotecount}}$ \marginpar{
\raggedright\tiny\em $\!\!\!\!\!\!\,\bullet$\themnotecount: #1} }

\newcommand{\alp}{{\alpha_p}}

\newcommand{\avtdpqk}{AVTD$^{(P,Q)}_k$}

\newcommand{\avtdpqtwo}{AVTD$^{(P,Q)}_2$}

\newcommand{\avtdpqi}{AVTD$^{(P,Q)}_\infty$}

\newcommand{\clim}{{\textstyle \lim_{C^0_{t_0}(\psi)}}}

\newcommand{\myqed}{\hfill $\Box$\\ \medskip }

\newcommand{\rmnote}[1]{}

\def \Reel{\mathbb{R}}
\def \R {\Reel}

\def \Nat{\mathbb{N}}

\def \N {\Nat}

\newcommand{\bel}[1]{\begin{equation}\label{#1}}
\newcommand{\beal}[1]{\begin{eqnarray}\label{#1}}
\newcommand{\bea}{\begin{eqnarray}}
\newcommand{\bean}{\begin{eqnarray}\nonumber}
\newcommand{\beadl}[1]{\begin{deqarr}\label{#1}}
\newcommand{\eeadl}[1]{\arrlabel{#1}\end{deqarr}}
\newcommand{\eeal}[1]{\label{#1}\end{eqnarray}}
\newcommand{\eead}[1]{\end{deqarr}}
\newcommand{\eea}{\end{eqnarray}}
\newcommand{\beaa}{\begin{eqnarray*}}
\newcommand{\eeaa}{\end{eqnarray*}}

\newcommand{\be}{\begin{equation}}
\newcommand{\ee}{\end{equation}}

\newcommand{\eq}[1]{\eqref{#1}}
\newcommand{\Eq}[1]{Equation~(\ref{#1})}

\DeclareFontFamily{OT1}{rsfs}{}
\DeclareFontShape{OT1}{rsfs}{m}{n}{ <-7> rsfs5 <7-10> rsfs7 <10->
rsfs10}{} \DeclareMathAlphabet{\mycal}{OT1}{rsfs}{m}{n}

\newcommand \al {\alpha}

\newcommand \pa {\partial}

\newcommand \mcM {{\mycal M}}

\newcommand \CH {{\mycal H}}
\newcommand \mcH {\CH}
\newcommand \cH {{\cal H}}

\newcommand \mcO {{\mycal O}}

\newcommand \CL {{\mycal  L}}

\newtheorem{defi}{\sc Definition\rm}[section]

\newtheorem{Theorem}[defi]{\sc Theorem\rm}

\newtheorem{Proposition}[defi]{\sc Proposition\rm}

\theorembodyfont{\upshape}
\newtheorem{Remark}[defi]{\sc Remark\rm}

\theoremstyle{nonumberplain}
 \theoremsymbol{$\Box$}
\theorembodyfont{\upshape}
\newtheorem{proof}{\sc Proof:\rm}

\begin{document}

\title{Cauchy horizons in  Gowdy space times}

\author{
Piotr T. Chru\'sciel\thanks{Partially supported by a Polish
Research Committee grant  KBN 2 P03B 073 24; email
\protect\url{piotr@gargan.math.univ-tours.fr}, URL
\protect\url{www.phys.univ-tours.fr\~piotr}} \\  D\'epartement de Math\'ematiques\\
Facult\'e des Sciences\\ Parc de Grandmont\\ F37200 Tours, France
\\
\\
Kayll Lake\thanks{Supported by a grant from the Natural Sciences
and Engineering Research Council of Canada; email
\protect\url{lake@astro.queensu.ca} . }\\Department of Physics and
\\ Department of Mathematics and Statistics\\ Queen's University\\
Kingston, Ontario, Canada K7L 3N6 }

\maketitle
\begin{abstract}
 We analyse
exhaustively the structure of \emph{non-degenerate} Cauchy
horizons in Gowdy space-times, and we establish existence of a
large class of non-polarized Gowdy space-times with such horizons.

Added in proof: Our results here, together with deep new results
of H.~Ringstr\"om (talk at the Miami Waves conference, January
2004), establish strong  cosmic censorship in (toroidal) Gowdy
space-times.
\end{abstract}

\section{Introduction}\label{sec:intro}
 In 1981 Vince Moncrief pointed out the
interest of studying the Gowdy metrics as a toy model in
mathematical general relativity, and proved the fundamental global
existence result for those metrics~\cite{Moncrief:Gowdy}. He
developed approximate methods to study their dynamics, and
discovered the leading order behavior of a large class of
solutions of the associated equations~\cite{grubisic93}. Together
with Beverly Berger he initiated the numerical investigation of
those models~\cite{Berger:1993ff}. In spite of considerable effort
by many researchers, the global properties of those models are far
from being understood. It is a pleasure to dedicate to him this
contribution to the topic.

In Gowdy space times  the essential part of the Einstein equations
reduces to a nonlinear wave-map-type system of
equations~\cite{GowdyANoP} for a map $x$ from $(M,g_{\al\beta})$
to the hyperbolic plane $(\CH,h_{ab})$, where $M=\lbrack T, 0)
\times \rm{S}^1$ with the flat metric $g=-dt^{2}+d{\theta}^2$. The
solutions are critical points of the Lagrangean
\begin{equation}\label{lag}
    \CL\lbrack x \rbrack = \frac 12 \int_{M} t g^{\al\beta }h_{ab}\pa
    _{\al}x^a\pa_{\beta }x^b \, d\theta dt\;.
    \end{equation} It is sometimes convenient to use coordinates
  $P,Q\in \R$  on the hyperbolic plane,  in which the hyperbolic metric $h_{ab}$
  takes the form
 \begin{equation}\label{pq}
 h= dP^2 + e^{2P}dQ^2.
 \end{equation}
 Let $ X_{t}=\frac {\pa x}{\pa t}, X_{\theta}= \frac {\pa x}{\pa
 \theta}$, $D$ denote the Levi-Civita connection of $h_{ab}$,
 and $D_\theta\equiv \frac {D}{D\theta}:=D_{X_{\theta}}$, $D_t\equiv\frac {D}{D  t}:=D_{X_{t}}$.
 The Euler-Lagrange equations for (\ref{lag}) take the form
 \begin{equation}\label{euler}
   \frac {DX_{t}}{D  t} - \frac {DX_{\theta}}{D\theta} =-  \frac {X_{t}}{ t}
 \end{equation}
or, in coordinates,
$$ \Box x^{a}+ \Gamma ^{a}_{bc}\circ x \pa_{\mu}x^b \pa^\mu x^{c}=
-  \frac {\pa _{t}x^a}{t},$$ where the $\Gamma$'s are the
Christoffel symbols of $h_{ab}$, and $\Box =
\partial_t^2-\partial_\theta^2$.

As already mentioned, V.~Moncrief established global existence of
smooth solutions on $(-\infty,0)$ of the Cauchy problem for
\eq{euler} \cite {Moncrief:Gowdy}. This
implies~\cite{Moncrief:Gowdy,ChAnop} that a maximal globally
hyperbolic Gowdy space-time $(\mcM,g)$ with toroidal Cauchy
surfaces can be covered by a global coordinate system
$(t,\theta,x^a)\in (-\infty,0)\times S^1\times S^1\times S^1$, in
which the metric takes the following form\footnote{\label{fnogood}
The form~\eq{Gspt} has been claimed by Gowdy~\cite{GowdyANoP}
under the hypothesis of two commuting Killing vectors, with
$\mathbb T^3$ spatial topology. This is not quite correct as there
are a few further global constants involved~\cite{ChAnop} even in
the current case of vanishing twist. However those constants can
be eliminated after passing from the torus ${\mathbb T}^3$ to
its universal cover $\R^3$, or when working in local coordinates,
and this suffices for the local considerations of the proofs given
here.}
\bean g&=& e^{-\gamma/2}|t|^{-1/2}(-dt^2+d\theta^2) + |t|e^P
(dx^1)^2+ 2 |t|e^PQ \,dx^1dx^2
\\ && +
|t|\left(e^PQ^2+e^{-P}\right)(dx^2)^2\;, \eeal{Gspt} where the
function $\gamma$ solves the equations \bel{Gspt2}
\partial_t\gamma = -t\left(|X_t|^2+|X_\theta|^2\right)\;,\quad
\partial_\theta \gamma = -2h(tX_t,X_\theta)\;.\ee
The main question of interest is the curvature blow-up -- or lack
thereof -- at the boundary $t=0$ of the associated space-time. An
exhaustive analysis of this has been carried out in~\cite{CIM} for
the so-called \emph{polarised case}, where the image of the map
$x$ is contained in a geodesic in the hyperbolic space. There
exist only two results in the literature which  prove curvature
blow up without the polarisation condition in a Cauchy problem
context: The first is due to one of the current authors (PTC), who
proves uniform curvature blow-up~\cite{SCC} under the condition
that the solution at $t=t_0$ satisfies\footnote{The threshold
${6^{-3/2} }$ in \eq{stcond} has been recently raised to $1/2$
in~\cite{Ringstroem4,ChCh2}.}
\bel{stcond}\sup_{\theta} t_0^2(|X_t|^2+|X_\theta|^2)(t_0,\theta)<
\frac 1{6^{3/2} }\;.\ee The second one\footnote{There is no
explicit statement about curvature blow-up in~\cite{Ringstroem3};
however, this follows immediately from the results
in~\cite{Ringstroem3} together with the calculations in the proof
of Theorem~3.5.1 of \cite{SCC}. The results in~\cite{Ringstroem3}
have been strengthened and generalised in~\cite{Ringstroem4}.} is
due to Ringstr\"om~\cite{Ringstroem3}, who assumes, at $t=t_0$,
smallness of the derivatives of $Q$ together with a bound
\bel{stcond2} 0<\eta \le |t_0|P_t(t_0,\theta)\le 1-\eta \;,\ee for
some strictly positive constant $\eta$. He further requires
$|t_0|$ to be sufficiently small.

The purpose of this paper is to study Cauchy horizons in Gowdy
space-times. Recall that existence of Cauchy horizons is precisely
what one wants to avoid in order to maintain predictability of the
Einstein equations. We wish therefore to describe properties of
those Gowdy space-times which possess Cauchy horizons, as a step
towards proving that generic initial data for those space-times
will \emph{not} lead to formation of Cauchy horizons. Our results
here, together with those
in~\cite{ChCh2,RendallWeaver,Ringstroem3,Ringstroem4,KichenassamyRendall,Rendall:2000ih},
give strong indications that this will be the case, though no
definitive statement is available so far.

 To put our results in proper context, we start by recalling
some results from~\cite{SCC} concerning the geometric properties
of the associated maximal globally hyperbolic solution
$(\mcM,{}^4g)$ of the Einstein equations. The following gives a
meaning to the statement that the set $\{t=0\}$ can be thought as
a (perhaps singular) boundary of the space-time:

\begin{Proposition}\label{Pinex}
Every future directed inextendible causal curve $(a,b)\ni s\to
\Gamma(s)$ in $(\mcM,{}^4g)$ reaches the boundary $t=0$ in finite
proper time. Further the limit $$\lim_{s\to
b}\theta(\Gamma(s))\;,$$ where $\theta$ is the coordinate
appearing in \eq{Gspt}, exists.
\end{Proposition}

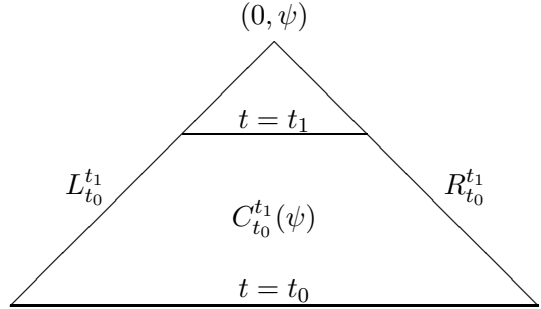
\begin{figure}
\begin{center}
\begin{picture}(150,120)(30,-10)
\thinlines \put(65,65){\line(1,0){70}} \put(0,0){\line(1,0){200}}
\put(0,0){\line(1,1){100}}
\put(200,0){\line(-1,1){100}}
\put(100,115){\makebox(0,0)[t]{$(0,\psi)$}}
\put(100,10){\makebox(0,0)[t]{$t=t_0$}}
\put(100,33){\makebox(0,0)[c]{$C^{t_1}_{t_0}(\psi)$}}
\put(164,45){\makebox(0,0)[l]{$R^{t_1}_{t_0}$}}
\put(36,45){\makebox(0,0)[r]{$L^{t_1}_{t_0}$}}
\put(100,70){\makebox(0,0)[c]{$t=t_1$}}
\end{picture}
\end{center}
\caption[PDE.1]{The truncated domains of dependence
$C^t_{t_0}(\psi)$.} \label{PDE.1}
\end{figure}

In order to continue, we need some terminology. For $t_0< 0$ let
the set $C^0_{t_0}(\psi)$ be defined as (compare
Figure~\ref{PDE.1})
\bel{Ctp} C^0_{t_0}(\psi)=\{t_0 \le t <0\;, \ - |t| \le
\theta - \psi \le |t|\}\;.\ee
We shall say that $\clim f = \alpha$
if
\bel{Cconv} \lim_{t\to0}\sup_{ - |t| \le \theta - \psi \le |t|}
|f(t,\theta)-\alpha| =0\;.\ee Such limits look a little awkward at
first sight; however, they arise naturally when considering the
behavior of the geometry along causal curves with endpoints on the
boundary $t=0$. In any case, the existence of such limits can
often be established in the problem at hand~\cite{SCC,ChCh2}.

The following relates properties of the Gowdy map $x$ to curvature
blow-up in space-time (the result follows immediately from the
arguments in the proofs of Theorem~3.5.1 and Proposition~3.5.2 in
\cite{SCC}):

\begin{Proposition}\label{Pcurvb}
Let $\psi\in S^1$ be such that \bel{derde}\clim |t^2D_\theta
X_\theta| =\clim |t^2D_\theta X_t| =\clim |tX_\theta| =0\;.\ee
  If there exists $1\ne v(\psi)\in \R$
such that \bel{derde2} \clim |tX_t| = v(\psi)\;,\ee then the
curvature scalar
$C_{\alpha\beta\gamma\delta}C^{\alpha\beta\gamma\delta}$ blows up
on every inextendible causal curve in $(\mcM,{}^4g)$ with end
point in $\{0\}\times \{\psi\}\times S^1\times S^1$.
\end{Proposition}

\medskip

\begin{Remark}
\label{Rcurvb} Condition \eq{derde2} can be weakened to
$$\mbox{\em either } \ \limsup |tX_t| <1 \; \ \mbox{\em or } \ \liminf |tX_t| > 1 \;,$$
where the limits sup and inf are understood  in a way analogous to
\eq{Cconv}.
\end{Remark}
\begin{Remark}Suppose that there exists a sequence of points
$(t_i,\theta_i)\in C^0_{t_0}(\psi)$ with $t_i \to 0$ and
$\Big||tX_t|(t_i,\theta_i)-1 \Big|>\epsilon$ for some $\epsilon$.
If \eq{derde} holds along the same sequence, one then obtains
curvature blow up along this sequence, which suffices to obtain
inextendibility of the space-time at $(0,\psi)$. From this point
of view the essential conditions are thus \eq{derde}, while the
existence of the velocity function $v(\psi)$ defined by
\eq{derde2} is actually irrelevant in the following sense: $v$
matters only at points at which it exists \emph{and} equals one.
\end{Remark}

Recall, now, that we are interested in conditions which would
guarantee that Cauchy horizons \emph{do not} occur. Consider,
thus, a Gowdy space-time with a Cauchy horizon $\cH$. It is then
easily seen (see Proposition~\ref{Pbpoint} below) that there
exists a point $\psi$ such that all curvature invariants remain
bounded along the timelike curves $t\to (t,\psi, x^1,x^2)$.
Proposition~\ref{Pcurvb} suggests then that the asymptotic
velocity $v$ should equal one at $\psi$. This is, however, not
entirely clear, because of the additional hypotheses \eq{derde}
made. Now, one would like to have a sharp Cauchy horizon criterion
without any unjustified restrictions. Here we prove the following
related statement:

\begin{Proposition}\label{Pveloc}
Consider a maximal globally hyperbolic Gowdy space-time which is
smoothly extendible across a Cauchy horizon $\cH$.
 Then there exists a point $\psi\in S^1$, coordinates $(P,Q)$ on the hyperbolic space,
 and a number $P_\infty(\psi)$
  such that
 \bel{veloc} \clim \left(P(t,\psi) +\ln
 |t|\right) = P_\infty(\psi)\;, \quad  \clim Q =0\;.\ee Further there exists a
 sequence $t_i\to 0$
such that
\bel{veloc2}\left(|t|\partial_t P\right)(t_i,\psi)\to 1\;.\ee
\end{Proposition}

To continue, one would like to understand precisely the geometry
of the set across which extensions are possible. Polarised
solutions are known which have $v=1$ on an interval
$[\theta_l,\theta_r]$, and which are extendible across
$\{0\}\times (\theta_l,\theta_r)\times S^1\times S^1$ ({\emph cf.,
e.g.,}\/~\cite{ChImaxTaubNUT}). This begs the question of
existence of non-polarised solutions with $v=1$ on an interval
$[\theta_l,\theta_r]$. In Section~\ref{Sv0} we provide a simple
construction of non-polarised Gowdy metrics, extendible across
such a smooth Cauchy horizon; this is rather similar to a
construction of Moncrief~\cite{MR88j:83027}
(compare~\cite{Berger:2000uf}).

Based on those examples, one would naively expect that
extendibility always requires the existence of such an interval
$[\theta_l,\theta_r]$. Further, in those examples, one has much
better control of the geometry than \eq{derde}; one has \avtdpqi\
behavior near $\{0\}\times [\theta_l,\theta_r]$, as defined at the
beginning of Section~\ref{S3}. Again one could hope that \avtdpqi\
behavior always holds across Cauchy horizons, but this has not
been proved so far. The main result of our paper is the following
theorem, which partially settles the issues raised so far:

\begin{Theorem}
\label{Thor} Consider a smooth Gowdy space-time that is extendible
across a Cauchy horizon $\cH $. Then:
\begin{enumerate}
\item $\cH $ is a Killing horizon; \emph{i.e.}, there exists a
Killing vector field which is tangent to the null geodesics
threading $\cH$.
\item \label{Thor2} $\cH$ is non-degenerate if and only if there
exists an interval $[\theta_l,\theta_r]\subset [a,b]$ such that
for $\theta\in [\theta_l,\theta_r]$ the velocity function
$v(\theta)$ exists and is equal to $1$ there, with
\bel{gvelo} |tP_t- 1| + |tP_\theta|+| te^PQ_t|+|te^PQ_\theta| \le
Ct^\alp\;,\ee for some $C,\alp>0$, in a certain coordinate system
$(P,Q)$ on hyperbolic space.
\item\label{Thor3} In fact, when $\cH$ is non-degenerate there
exists a $(P,Q)$ coordinate system on the hyperbolic space $\mcH
_2$ in which the solution is \avtdpqi\ near $\{0\}\times
[\theta_l,\theta_r]$, with $Q_\infty=0$ over
$[\theta_l,\theta_r]$, and with the conclusions
of Theorem~\ref{Tsvel2} below holding on $[\theta_l,\theta_r]$. 
\end{enumerate}
\end{Theorem}

\begin{Remark} Assuming smooth Cauchy data, we note that there exists $k_0\in \N$ such
that if the metric is $C^{k_0}$ extendible across a non-degenerate
$\cH $, then it is smoothly extendible there. An explicit estimate
for $k_0$ can be found by chasing differentiability in the
constructions in the proof below to ensure that \eq{svel12} holds.
Smoothness of the extension follows then from
Theorem~\ref{Tsvel2}. Similar results can be established for Gowdy
metrics with (sufficiently high) finite differentiability, whether
in a classical or in a Sobolev sense.
\end{Remark}

\begin{Remark}
Extendibility above is meant in the class of Lorentzian manifolds;
no field equations are assumed to be satisfied by the extension.
We note that a necessary condition, in the non-degenerate case,
for extendibility in the class of vacuum metrics is analyticity of
the metric.
\end{Remark}

 Theorem~\ref{Thor} shows that degenerate
 horizons $\cH$, if any,
 would correspond to a single point in the $(t,\theta)$ coordinates.
 For analytic metrics non-existence of  degenerate horizons $\cH $ in the Gowdy
 class of metrics
 follows from the arguments of the proof of Theorem~2
of~\cite{VinceJimcompactCauchyCMP}, but the information provided
by those considerations  does not seem to suffice to exclude the
possibility that $\kappa=0$ for smooth but not necessarily
analytic metrics. While we find it unlikely that degenerate horizons
could exist here, we can exclude them only under some
supplementary hypotheses on the behavior of the Gowdy map
$x$:

\begin{Proposition}
\label{Pcurv2} Let $\psi$ be such that $v(\psi)=1$ and suppose
that either
\begin{enumerate}
\item[a)] there exists $\alp>0$ such that  we have\bel{svel12b} |t\partial_\theta P|+ |te^P \partial _t
Q|\le C|t|^\alp\ee  on $C^0_{t_0}(\psi)$, or
\item[b)] \Eq{derde} holds, or
\item[c)] the solution is \avtdpqtwo\ on $C^0_{t_0}(\psi)$.
\end{enumerate}
Then: \begin{enumerate}\item  
If there exists a sequence  $\psi_i\to\psi$ such that
$v(\psi_i)\ne 1$, 
then there exists no extension of $\mcM$ across
$\{0\}\times{\psi}\times S^1\times S^1$. \item Moreover, if
$$- \left( {\frac {d}{d\theta}}v
\left( \psi
 \right)  \right) ^{2}+{e^{2\,P_{{\infty }} \left( \psi \right) }}
 \left( {\frac {d^{2}}{d{\theta}^{2}}}Q_{{\infty }} \left( \psi
 \right)  \right) ^{2} \ne 0 \;,$$ then the curvature scalar
$C_{\alpha\beta\gamma\delta}C^{\alpha\beta\gamma\delta}$ blows up
on every inextendible causal curve in $(\mcM,{}^4g)$ with
accumulation point in $\{0\}\times \{\psi\}\times S^1\times S^1$.
\end{enumerate}
\end{Proposition}

We note that the power-law condition~\eq{svel12b} is justified for
initial data near those corresponding to a flat Kasner solution
$(P_0,Q_0)\equiv ( -\ln |t|, 0)$ by the results in~\cite{ChCh2}.

 This paper is organised as follows: In
Section~\ref{Sv0} we present our construction of solutions with
Cauchy horizons. In Section~\ref{S3} we define \avtdpqk\ Gowdy
maps, and we give proofs of the results described above. In
Section~\ref{Scurv} we investigate further the blow-up of the
Kretschmann scalar
$C_{\alpha\beta\gamma\delta}C^{\alpha\beta\gamma\delta}$. In
Appendix~\ref{SA} the Riemann tensor of the Gowdy metrics is
given.

\medskip

{\noindent \bf Added in proof:} It has been shown recently by
H.~Ringstr\"om~\cite{Ringstroem4} that \eq{derde} holds at all
points $\theta\in S^1$ for all solutions of Gowdy equations. The
arguments of H.~Ringstr\"om~\cite{Ringstroem4} also show that the
collection of Gowdy initial data for which the set $\{\theta:
v(\theta)=\pm 1\}$ has no interior is of second category. Consider
any solution in that last class, and suppose it contains a Cauchy
horizon. Proposition~\ref{Pcurv2} shows that the horizon is
non-degenerate, which is impossible by Theorem~\ref{Thor}. It
follows that generic (in the second category sense) Gowdy metrics
on $T^3$ have no Cauchy horizons, which establishes strong cosmic
censorship within this class of metrics.

\section{Two families  of solutions with $v\equiv 0$ and with $v\equiv 1$}
\label{Sv0}

 There are by now at least two systematic ways
~\cite{Rendall:2000ih,Ringstroem3} of constructing reasonably
general families of  solutions with controlled asymptotic behavior
satisfying (compare \eq{stcond2}) \bel{stcond22} 0<\eta \le
|t|P_t(t,\theta)\le 1-\eta\qquad \forall \ \theta \in S^1\;,\
t_0\le t<0 \;.\ee The proofs given there only cover situations in
which\footnote{However, it seems that certain families of
solutions with vanishing $v$ can also be constructed using
Fuchsian techniques (A.~Rendall private communication).} $|t|P_t$
is bounded away from zero. It turns out that there is a trivial
way of constructing families of solutions for which $tX_t$
approaches zero asymptotically, as follows: Let
$$y\left(x^0,r=\sqrt{(x^1)^2+(x^2)^2}\right)$$ be any
rotationally-symmetric solution of the standard wave-map equation
from the three-dimensional Minkowski space-time $(\R^{1,2},\eta)$
into the two-dimensional hyperbolic space $(\mcH_2,h)$.  Set
\bel{zvs} x(t,\theta) = y(x^0=\theta,r=-t)\;.\ee Note that the
replacement $|t|\to r$, $\theta\to x^0$ brings the action \eq{lag}
to that for a rotationally invariant wave-map as above, so it
should be clear that
 $x$ satisfies the
Gowdy evolution equation~\eq{euler}. Now, the solution $x$ will
not be periodic in $\theta$ in general (and it seems that there
are no non-trivial smooth wave maps $y$ at all which are periodic
in $x^0$). However, it is straightforward to construct periodic
solutions for which \eq{zvs} will hold on an open neighborhood of
some interval $\{0\}\times [b,2\pi -b]\subset \{0\}\times
[0,2\pi]\approx \{0\}\times S^1$, with $0<b<\pi$, as follows: let
$x$ be as in \eq{zvs}, and for $\theta\in [b/2,2\pi-b/2]$ let
$$f(\theta)=x(-b/2,\theta)\;,\quad g(\theta)= \partial_t
x(-b/2,\theta)\;.$$ Extend $f$ and $g$ to $2\pi$--periodic
functions $(\hat f,\hat g)$ in any way. Let $x_{\hat f,\hat g}$ be
the solution of the Gowdy evolution equation~\eq{euler} with
initial data $(\hat f,\hat g)$ at $t=-b/2$. Uniqueness in domains
of dependence of solutions of \eq{euler} shows that $x_{\hat
f,\hat g}$ will coincide with $x$ in the domain of dependence of
the set $\{t=-b/2\,,\ \theta\in [b/2,2\pi-b/2]\}$:
$$x_{\hat f,\hat g}=x \ \mbox{ on } \  \{-b/2 \le t \le 0\,,\ b+t \le \theta \le
2\pi- b-t\}\;.$$ Clearly one can also construct solutions which
will satisfy \eq{zvs} on the  union of any finite number of
disjoint domains of dependence as above.

 The wave maps $y$ can be obtained
by prescribing arbitrary rotationally invariant, say smooth,
Cauchy data $y(0,r)$ and $\partial_t y(0,r)$, and the solutions
are always global~\cite{CT93}. In this way one obtains a family of
solutions $x$ of the Gowdy equations parameterised, locally, by
four free smooth functions (this, by the way, shows that the
``function counting method" might be rather misleading, as this
family of solutions does certainly {\em not} form an open set in
the set of all solutions). The $y$'s are globally smooth both in
the $t$ and $\theta$ variables, which implies that the $x$'s
display the usual ``asymptotically velocity dominated" behavior
with \bel{zerv}\lim_{t\to 0}tX_t = \lim_{t\to 0}X_t=0\;,\ee the
convergence being uniform in $\theta$ for $\theta\in[a,b]$. In
particular one has curvature blow-up for causal curves ending on
$\{0\}\times [a,b]\times S^1\times S^1$ by
Proposition~\ref{Pcurvb}.

 \Eq{zerv} shows
that $x$ has zero velocity in the relevant range of $\theta$'s. It
follows from the results proved in~\cite{ChCh2} that all solutions
which have $v\equiv 0$ on an interval $I$ are obtained from the
procedure above on an neighborhood of $\{0\}\times I$.

 The solutions just described can be used to
construct solutions with $v=1$ on open intervals. (Here and
elsewhere $v$ is always the quantity defined as in \eq{derde2},
whenever it exists, with $X_t$ being associated with the $(P,Q)$
representation of the solution as in \eq{Gspt}.) Indeed, let
$x=(P,Q)$ be one of the zero-velocity solutions defined by
\eq{zvs}, and define a new solution $\hat x$ by performing the
``Gowdy-to-Ernst" transformation~\cite{RendallWeaver}: \bel{GtE}
\hat P := -P - \ln |t|\;, \quad e^{\hat P} \partial _t \hat Q: = -
e^P \partial _\theta Q\;, \quad e^{\hat P} \partial _\theta \hat
Q: = - e^P
\partial_t Q\;. \ee The new map satisfies again the Gowdy
equation~\eq{euler}. It immediately follows that we have
$$\Big| |tX_t|-1\Big| + |tX_\theta| \le C|t|\;,$$
so that we obtain power-law blow-up, together with the new
asymptotic velocity equal to one on any interval on which the old
velocity $v=0$. All the resulting space-times have a smooth Cauchy
horizon across any interval on which  $v=\pm1$, which can be
checked by standard calculations using \eq{sh46a}
(compare~\cite{ChImaxTaubNUT,MR88j:83027,MR86c:83009,MR82b:83024}).

As already mentioned in the introduction, a similar technique,
mapping singular solutions to ones with Cauchy horizons and
vice-versa, has been used by V.~Moncrief in~\cite{MR88j:83027},
compare~\cite{Berger:2000uf}. While V.~Moncrief used this method
to produce singular solutions out of ones with Cauchy horizons,
our approach is exactly the reverse one.

\section{Proofs} 
\label{S3}

We start with some terminology. We will be mainly interested in
solutions of \eq{euler} with  the following behavior (for
justification,
see~\cite{CIM,KichenassamyRendall,Rendall:2000ih,ChCh2} and
\eq{Pexp}-\eq{Qexp} below):
\beal{avtd1} P(t,\theta) &=& -v_1(\theta)\ln |t| +P_\infty(\theta)
+ o(1)\;,
\\ Q(t,\theta) &=& Q_\infty(\theta) + \left\{%
\begin{array}{ll}
    |t|^{2v_1(\theta)}\Big(\psi_Q(\theta) + o(1)\Big)\;, & \hbox{$0< v_1(\theta)\not\in \N$ ;} \\
    |t|^{2v_1(\theta)}\Big(Q_{\ln}(\theta)\ln |t|+\psi_Q(\theta) + o(1)\Big)\;, & \hbox{$0<v_1(\theta)\in \N$ ;} \\
    Q_{\ln}(\theta)\ln |t| + o(1)\;, & \hbox{$ v_1(\theta)\in -\N^*$ ;} \\
    o(1)\;, & \hbox{$-\N^*\not\ni v_1(\theta)\le 0$ .}  
\end{array}%
\right.\nonumber \\&& \eeal{avtd2} The function \bel{vdefn}
v:=|v_1|\ee will be called the \emph{velocity function}, while
$Q_\infty$ will be called the
\emph{position function}. 
We shall say that a solution is in the \avtdpqk\ class if
\eq{avtd1}-\eq{avtd2} hold with functions $v_1$, $P_\infty$,
$Q_\infty$ and $\psi_Q$  which are of $C_k$ differentiability
class (on closed intervals the derivatives are understood as
one-sided ones at the end points). For $k>0$ we will assume that
the behavior \eq{avtd1}-\eq{avtd2} is preserved under
differentiation in the following way: \bel{avtd1d} \forall\ 0\le
i+j\le k \qquad
\partial_\theta^j(t\partial_t)^i\Big(P(t,\theta)
+v_1(\theta)\ln |t| -P_\infty(\theta)\Big) = o(1)\;,\ee similarly
for $Q$.

\medskip

We continue with the

\medskip

\noindent{\sc Proof  of Proposition~\ref{Pinex}:} The fact that
inextendible causal curves meet the set $t=0$ in finite time is
proved in Proposition~3.5.1 in~\cite{SCC}. Next, since $t$ is a
time function on $\mcM$, we can parameterize $\Gamma$ by $t$:
$$\Gamma(t)=(t,\theta(t),x^a(t))\;.$$
Timelikeness of $\Gamma$ together with the form \eq{Gspt} of the
metric gives
$$\left|\frac{d\theta}{dt}\right| \le 1\;,$$
which implies that $\theta$ is uniformly Lipschitz along $\Gamma$,
and the existence of the limit $\psi:=\lim_{t\to 0}\theta(t)$
immediately follows.

\myqed

We are ready now to pass to the proof of the following result,
mentioned in the Introduction:

\begin{Proposition}
\label{Pbpoint}Consider a Gowdy space-time with a Cauchy horizon
$\cH$. Then there exists   $\psi\in S^1$ such that all curvature
invariants remain bounded along the timelike curves $t\to (t,\psi,
x^1,x^2)$, $-1\le t<0$.
\end{Proposition}

\proof  Let $p\in\cH$ and let $\hat \Gamma$ be any timelike curve
starting from $p$ and entering the globally hyperbolic region
$\mcM$. We set $\Gamma:=\hat \Gamma \cap \mcM$; by
Proposition~\ref{Pinex} there exists $\psi\in S^1$ such that
$\theta$ approaches $\psi$ along $\Gamma$. By hypothesis the
extended metric is smooth around $p$, hence there  exists a
neighborhood $\mcO$ of $p$ on which all curvature invariants are
bounded. In particular all curvature invariants are bounded on
$\mcO\cap \mcM$. Since the metric is $U(1)\times U(1)$ invariant
on $\mcM$,  all the curvature invariants will also be bounded on
the set obtained by moving $\mcO\cap \mcM$ by isometries. This
last set contains all the timelike curves as in the statement of
the Proposition, with $-\epsilon < t <0$, and the result easily
follows. \myqed

 One of the ingredients of the proof of Theorem~\ref{Thor}
is the following result, proved in~\cite{ChCh2}:
\begin{Theorem}
\label{Tsvel2}  Suppose that there exist constants  $C,\alp>0$
such that \bel{svel12} |t\partial_\theta P|+ |te^P \partial _t
Q|\le C|t|^\alp\ee and consider any point $\psi\in[a,b]$ such that
$v_1(\psi)=1$. Then the functions $(\bar P,\bar Q):= (P+\ln |t|,
Q)|_{C^0_{t_0}(\psi)}$ can be extended  to an \avtdpqi\ map from
$\R^2$ to $\mcH_2$. If\, $\clim t\partial^j_\theta P_t = 0$ for
all $j\in\N$, then for all $i,k\in\N$ we have
 \bel{sh46a} \clim \partial_t^{2i+1}\partial _\theta^k (P +\ln |t|)= \clim \partial_t^{2i+1}\partial _\theta^k Q
 =\clim \partial _\theta^{k+1} Q =0\;.\ee
Further, if $v_1=1$  on an interval $[\theta_l,\theta_r]$, then
the restriction $(\tilde P,\tilde Q):= (P+\ln |t|,
Q)|_{\Omega(\theta_l,\theta_r,t_0)}$  can be extended  to a smooth
map from $\R^2$ to $\mcH_2$, with \eq{sh46a} holding for all
$\psi\in[\theta_l,\theta_r]$.
\end{Theorem}

\begin{Remark} \label{Rsvel12} The vanishing of the last term in \eq{sh46a} for all $k\ge 0$ is
somewhat surprising. As already pointed out in the introduction,
the power-law condition~\eq{svel12} is justified for initial data
near those corresponding to a flat Kasner solution
$(P_0,Q_0)\equiv ( -\ln |t|, 0)$ by the results in~\cite{ChCh2}.
\end{Remark}

We are ready now to pass to the

\medskip

\noindent{\sc Proof of Theorem~\ref{Thor}:} Recall that $\cH $ is
a Killing horizon if there exists a Killing vector field which is
tangent to the generators of $\cH $. The fact that $\cH $ is a
Killing horizon follows from the
 proof of Proposition~1 of \cite{BCIM} (see
in particular Lemma~1.1 and Lemma~1.2 there); further, any other
Killing vector is spacelike on $\cH $. Thus, if a Gowdy space-time
is extendible across a connected Cauchy horizon, then one of the
Killing vectors, say $X$, is tangent to the generators of the
event horizon. Since our claims are local, without loss of
generality we may assume that $\cH $ is connected. Locally near a
point in $\cH $ we can construct a null Gauss coordinate system
$(T,W,\hat x^A)$, $A=1,2$, as in~\cite{VinceJimcompactCauchyCMP},
leading to the following local form of the metric \bel{ncs} g =
2dT\,dW + T\phi\, dW^2 + 2 T\beta_Ad\hat x^A dW + \mu_{AB}d\hat
x^A d\hat x^B\;,\ee with some smooth functions $\phi$, $\beta_A$,
and $\mu_{AB}$ which have obvious tensorial properties with
respect to $\hat x^A$--coordinate-transformations. More precisely,
let $(W,\hat x^A)$ be local coordinates on $\cH $ such that
$\partial_W=X$, and such that $\partial_2$ is another linearly
independent Killing vector. Let $k$ be the null vector field
defined along $\cH $, pointing towards the globally hyperbolic
region, such that
$$g(\partial_W,k)=1\;,\quad g(\partial_A,k)=0$$  on $\cH $.
The local coordinates $(W,\hat x^A)$ are extended from $\cH $ to a
neighborhood thereof by requiring  $(W,\hat x^A)$ to be constant
along the geodesics issued from $\cH $ with initial velocity $k$.
Letting $T$ be the affine parameter along those geodesics, with
$T=0$ on $\cH $, one obtains \eq{ncs}. Now, isometries map
geodesics to geodesics and preserve affine parameterisations,
which easily implies
$$\partial_W g_{\mu\nu}=0=\partial_2 g_{\mu\nu}\;,$$ throughout
the domain of definition of the coordinates. Equation~(2.9)
of~\cite{VinceJimcompactCauchyCMP} for $R_{3b}$ shows that on any
connected component of $\cH $ there exists a constant $\kappa\ge
0$ such that
$$\phi|_{\cH }=\kappa\;.$$

Whatever the range of the $W$ and $\hat x^2$ coordinate in the
original extension, we can without loss of generality assume that
the functions above are defined for $W, \hat x^2$ in $\R$ -- or in
$S^1$ -- since those functions are independent of $W$ and $\hat
x^2$ anyway. There might be difficulties if we are trying to build a
manifold by patching together the resulting local coordinates, but
we do not need to patch things back together, so this is
irrelevant for the local calculations that follow.

For $T>0$ we replace the coordinates $T$ and $W$ by new
coordinates $(\hat t,\hat x^3)$, $\hat t> 0$, defined as
\bel{ncs2} W = \hat x^3 + \alpha \ln \hat t \;,\quad T= \beta \hat
t^2\;,\ee with constants $\alpha\in \R$, $\beta> 0$, leading to
\bean g & =& \alpha\beta\left(4+\kappa \alpha+O(\hat
t^2)\right)d\hat t^2+ 2\beta\left(2+\alpha\kappa+O(\hat
t^2)\right)\hat t\,d\hat td\hat x^3
\\ && + 2 \hat t\beta_Ad\hat x^A (\hat td\hat x^3+ \alpha d\hat t)
+ \beta \hat t^2(\kappa+O(\hat t^2))(d\hat x^3)^2
\nonumber \\ &&+ \mu_{AB}d\hat x^A d\hat x^B\;.\eeal{ncs3} We can
choose $\alpha,\beta$ so that $$\alpha\beta(4+\kappa\alpha)
<0\;,$$ and we assume that some such choice has been made.

Let $X_1=\partial_W=\partial_{\hat x^3}$, $X_2=\partial_{\hat
x^2}$,  and define
$$\lambda_{ab}= g(X_a,X_b)\;,\quad a,b=1,2\;.$$
From \eq{ncs3} we have \bea 
& \lambda_{11}=  \beta\kappa
\hat t^2+O(\hat t^4)\;,\quad \lambda_{12}=  \beta_1 \hat
t^2+O(\hat t^4)\;, 
\quad  \lambda_{22}=\mu_{22}\;, &  \eeal{ncs5}%
Note that $X_2$ is spacelike on $\cH $.
It follows that
\bel{tek0}\partial_\mu \left(\sqrt{\det \lambda}\right) 
=\left\{%
\begin{array}{ll}
    \sqrt{\beta \kappa\mu_{22}}+O(\hat t^2), & \hbox{$\partial_\mu=\partial_{\hat t}$, $\kappa\ne 0$,} \\
    O(\hat t), & \hbox{$\partial_\mu=\partial_{\hat t}$, $\kappa= 0,$} \\
    \frac 12 \sqrt{\frac{\beta \kappa}{\mu_{22}}}\partial_{\hat x^1}\left(\mu_{22}\right) \hat t
    + O(\hat t^3), & \hbox{$\partial_\mu=\partial_{\hat x^1}$, $\kappa\ne 0,$} \\
    O(\hat t^2), & \hbox{$\partial_\mu=\partial_{\hat x^1}$, $\kappa= 0,$} \\
    0, & \hbox{$\partial_\mu=\partial_{\hat x^i}$, $i=2,3$,} \\
\end{array}%
\right.    \ee
 and \bel{teq} t:=-\sqrt{\det \lambda} =\left\{%
\begin{array}{ll}
    -\sqrt{\beta \kappa \mu_{22}}\hat t + O(\hat t^3), & \hbox{$\kappa\ne 0$,} \\
    O(\hat t^2), & \hbox{$\kappa= 0$.} \\
\end{array}%
\right. \ee Now,  away from the set $t=0$ the space-time metric
$g$ can be written$^{\mbox{\scriptsize \ref{fnogood}}}$ in the
form \eq{Gspt}. The functions $t$ and $\theta$ are defined
uniquely up to a single multiplicative constant, \emph{cf.,
e.g.,\/}~\cite{ChAnop}; the normalisation \eq{teq} gets rid of
that freedom. By a rotation of the Killing vectors
$\partial_{x^a}$ we can always achieve
$$\partial_{x^1}=X_1=\partial_{\hat x^3}\;,\quad \partial_{x^2}=X_2=\partial_{\hat x^2}\;.$$
Recall that a Killing horizon is said to be degenerate if
$\partial_\mu \left(g(X,X)\right)|_{\cH }=0$. Since $\mu_{22}$
does not vanish at $\hat t=0$, \eq{teq} shows that $\cH $ is
degenerate if and
only if 
$\kappa$ vanishes. From now on we assume that $\kappa\ne 0$.
Inspection of \eq{ncs3} shows that a convenient choice of $\alpha$
and $\beta$ is
$$2+\alpha\kappa=0\;,\quad \alpha\beta(4+\kappa\alpha) =-1\;,$$
so that $\beta=\kappa/4$. Ordering the entries as $(\hat t, \hat
x^3,\hat x^1,\hat x^2)$, \eq{ncs3} takes the following matrix
form  $$g=\left(%
\begin{array}{cccc}
 -1+O(\hat t^2) & O(\hat t^3) & O(\hat t) & O(\hat t) \\
  O(\hat t^3) & \beta\kappa \hat t^2 +O(\hat t^4) & O(\hat t^2) & O(\hat t^2) \\
  O(\hat t) & O(\hat t^2) & \mu_{11} & \mu_{12} \\
  O(\hat t) & O(\hat t^2) & \mu_{12} & \mu_{22} \\
\end{array}%
\right) \;.$$ This gives $\det g = -\beta\kappa \hat t^2\det \mu +
O(\hat t^4)$ and \bel{ncs20}g^{-1}=\left(%
\begin{array}{cccc}
 -1+O(\hat t^2) & O(\hat t) & O(\hat t) & O(\hat t) \\
  O(\hat t) & (\beta\kappa \hat t^2)^{-1} +O(1) & O(1) & O(1) \\
  O(\hat t) & O(1) & \mu^{11} +O(\hat t^2) & \mu^{12} +O(\hat t^2)\\
  O(\hat t) & O(1) & \mu^{12}+O(\hat t^2) & \mu^{22}+O(\hat t^2) \\
\end{array}%
\right) \;,\ee where $\mu^{AB}$ is the matrix inverse to
$\mu_{AB}$. From \eq{ncs20} and \eq{tek0} we find for $t<0$
\bean g(\nabla t,\nabla t)  &= & g^{\hat t \hat t} \left(\frac{\partial t}{\partial \hat t}\right)^2 +
2g^{\hat t \hat x^1} \frac{\partial t}{\partial \hat
t}\frac{\partial t}{\partial \hat x^1} + g^{\hat x^1 \hat x^1}
\left(\frac{\partial t}{\partial \hat x^1}\right)^2
\\&= & \left(-1+O(\hat t^2)\right) \left(\frac{\partial t}{\partial \hat t}\right)^2 +
O(\hat t) \frac{\partial t}{\partial \hat t}\frac{\partial
t}{\partial \hat x^1} + \left(\mu^{11}+O(\hat
t^2)\right)\left(\frac{\partial t}{\partial \hat
x^1}\right)^2\nonumber
\\ & = &
 -\beta\kappa\mu_{22}
 +O(\hat t^2)\;.\eeal{ncs18}
Comparing \eq{Gspt} and \eq{ncs20} we also obtain
$$0=g(\nabla t,\nabla\theta)= -\left(\sqrt{\beta \kappa \mu_{22}}+O(\hat t^2)\right)\frac{\partial
\theta}{\partial \hat t} +
O(\hat t)
\frac{\partial \theta}{\partial \hat x^1} \;,$$ and \beaa
\beta\kappa\mu_{22}
 +O(\hat t^2)& =&-g(\nabla t,\nabla t)= g(\nabla
\theta,\nabla\theta)\\  &= & g^{\hat t \hat t}
\left(\frac{\partial \theta}{\partial \hat t}\right)^2 + 2g^{\hat
t \hat x^1} \frac{\partial \theta}{\partial \hat t}\frac{\partial
\theta}{\partial \hat x^1} + g^{\hat x^1 \hat x^1}
\left(\frac{\partial \theta}{\partial \hat x^1}\right)^2
\\
&= & -\left(1 + O(\hat t^2)\right)\left(\frac{\partial
\theta}{\partial \hat t} \right)^2 + O(\hat t)\frac{\partial
\theta}{\partial \hat t}\frac{\partial \theta}{\partial \hat x^1}
\\ && +\left(
\mu^{11}+O(\hat t)\right)\left(\frac{\partial \theta}{\partial
\hat x^1} \right)^2 \;.\eeaa It follows that ${\partial
\theta}/{\partial \hat x^1}$ is uniformly bounded from above and
away from zero for $t>0$, with $\partial_{\hat t}\theta = O(\hat
t)$. This implies that $\theta$ extends by continuity to a
Lipschitz function on $\cH $, and also implies the existence of the
claimed interval of $\theta$'s.

Comparing \eq{ncs5} with \eq{Gspt} we find
$$ |t|e^P = \beta \kappa\hat t^2 +   O(\hat t^4)\;,\quad |t|e^P Q = \beta_1 \hat t^2 +   O(\hat t^4)\;.$$
For $\kappa \ne 0 $ this gives
$$ e^P = \sqrt{\frac{\beta \kappa}{ \mu_{22}}}\hat t +   O(\hat t^3)= \frac{| t|}{\mu_{22}} + O(|t|^3)\;,
\quad Q = \frac{\beta_1}{\beta\kappa}  +   O(\hat t^2)\;,$$ and
what has been shown so far about $t$ and $\theta$ further gives
$$tP_t= 1 + O(t^2)\;,\ tP_\theta = O(|t|)\;,\ te^PQ_t = O(|t|^3)\;,\ te^PQ_\theta =
O(|t|^2)\;.
$$
We have thus obtained a representation of the solution for which
\bel{goodbeh} v_1=-1\;,\quad v=1\;,\quad |tX_\theta| =
O(|t|^3)\;,\ee and the sufficiency part of point~(ii) is
established. At this stage one can directly derive a full
asymptotic expansion of the solution using the Gowdy
equations~\eq{euler}; this will lead to \avtdpqi\ behavior with a
perhaps non-constant function $Q_\infty$. An alternative way
consists in swapping the order of the Killing vectors;
\eq{goodbeh} will then still hold except for a change of sign of
$v_1$, so  that the hypotheses of Theorem~\ref{Tsvel2} are
satisfied, and point~(iii) follows.

Consider, finally, a solution with $v=1$ at a point $\psi$, or on
an interval $[\theta_l,\theta_r]$, with \eq{gvelo} holding there.
By Theorem~\ref{Tsvel2} the solution is \avtdpqi\ on
$C^0_{t_0}(\psi)$, or on $[t_0,0)\times [\theta_l,\theta_r]$, and
by integration of \eq{Gspt2} one easily finds
\bel{gbeh}\gamma=-\ln|t|+O(1)\;.\ee As already pointed out, such solutions are
smoothly extendible by the calculations in \cite{ChImaxTaubNUT};
alternatively, one can run backwards the calculations done above. In
any case  \eq{ncs18} holds. From \eq{Gspt} we have
$$g(\nabla t,\nabla t) = -
e^{\gamma/2}|t|^{1/2}\;,$$ and comparing \eq{gbeh} with \eq{ncs18}
the non-degeneracy of the horizon follows.

 \myqed

 The reader will have noticed that the last part of the proof of
 Theorem~\ref{Thor} shows the following:

 \begin{Proposition}
 \label{Lnodeg} \avtdpqi\ space-times do not contain degenerate
 horizons.
 \end{Proposition}

 We continue with the

 \medskip

\noindent{\sc Proof of Proposition~\ref{Pveloc}:} The result
follows immediately from the calculations in the proof of
Theorem~\ref{Thor}, with the following modifications: we choose
$X_2$ to be the Killing vector which is tangent to the generators
of $\cH$, and we let $X_1$ be any other Killing vector. If
$\psi$ is a point as in Proposition~\ref{Pbpoint}, then
$g(X_1,X_1)$ tends to a strictly positive  limit along the curve
$\Gamma$ defined in the proof of Proposition~\ref{Pbpoint}.
Further this limit is the same for any causal curve which
accumulates at the point $p$ of the proof of
Proposition~\ref{Pbpoint}.  It follows that along any of the
curves $t\to (t,\psi,x^a)$ the limit $\lim_{t\to 0} g(X_1,X_1)$
exists, and in fact one has that the limit $$\clim g(X_1,X_1)$$
exists. Using \eq{Gspt}, this can be translated into the statement
that the function $te^{P(t,\psi)}$ has a finite limit as $t$ goes
to zero, which justifies the first equation in \eq{veloc}. Since
$X_2$ is normal to $\cH$ the function $g(X_1,X_2)$ vanishes on
$\cH$, which implies the second equation in \eq{veloc}. Finally if
$|t|\partial_tP$ avoids the value one, then either
$|t|\partial_tP<1-\epsilon$ or $|t|\partial_tP>1+\epsilon$ for
some $\epsilon>0$ for $t$ large enough, which is incompatible with
the first equation in \eq{veloc}. \myqed

We close this section with the

\medskip

\noindent{\sc Proof of Proposition~\ref{Pcurv2}:} Without loss of
generality we can choose the coordinates $(P,Q)$ on hyperbolic
space so that $v_1(\psi)=1$. It follows from the results
in~\cite{ChCh2} that under any of the conditions of
Proposition~\ref{Pcurv2} the conclusions of Theorem~\ref{Tsvel2}
hold; thus $x$ is \avtdpqi\ in $C^0_{t_0}(\psi)$. Under the
hypotheses of point (i), suppose that the solution is extendible.
Since the velocity function is not equal to one in an open
interval
 containing $\psi$,  Theorem~\ref{Thor} shows that the horizon must be degenerate.
 This contradicts Proposition~\ref{Lnodeg}, and establishes point
 (i).

 Point (ii) can then be established by
inspection of the curvature tensor, which we give for completeness
in Appendix~\ref{SA}. The relevant calculations have been done
using Grtensor~\cite{Ishak:2001uj}. The interested reader will
find MAPLE worksheets and input files on URL
\url{http://grtensor.phy.queensu.ca/gowdy}. It follows from the
the \avtdpqi\ character of the solution on $C^0_{t_0}(\psi)$ that
there exist bounded functions $\gamma_\infty$ and $Z$ such that
$$\gamma = -\ln |t| +  \gamma_\infty(\theta) + o(1)\;,$$
$$P(t,\theta)=-v(\theta)\ln|t|+P_\infty(\theta)+tZ(t,\theta)\;.$$
The field equations further show that $\lim_{t\to 0}t \partial_t
Z(t,\psi)=0$ (in fact $Z= O(|t|\ln |t|) $; see the next section
for more detailed expansions), and one finds
\begin{equation}
C_{\alpha \beta \gamma \delta}C^{\alpha \beta \gamma \delta} \sim
4\,{\frac { \left( - \left( {\frac {d}{d\theta}}v \left( \psi
 \right)  \right) ^{2}+{e^{2\,P_{{\infty }} \left( \psi \right) }}
 \left( {\frac {d^{2}}{d{\theta}^{2}}}Q_{{\infty }} \left( \psi
 \right)  \right) ^{2} \right) {e^{\gamma \left( t,\psi \right) }}}{
t}}\;, \label{basiccc}
\end{equation}
as $t \rightarrow 0$, provided that the coefficient in the biggest
parenthesis in the  numerator does not vanish. As $e^{\gamma(
t,\theta)} \sim 1/|t|$ the curvature scalar $C_{\alpha
\beta \gamma \delta}C^{\alpha
\beta \gamma \delta}$ diverges then like $1/t^2$. (As already
pointed out in the introduction, in the next section we will study
in detail the remaining possible behaviors of $C_{\alpha
\beta \gamma \delta}C^{\alpha
\beta \gamma}$.) \myqed

\section{The blow-up structure of the Kretschmann
scalar}\label{Scurv}

Throughout this section we assume that $\psi$ is such that
$v_1(\psi)=1$. To analyse the behavior of $C_{\alpha
\beta \gamma \delta}C^{\alpha
\beta \gamma \delta}$ in the case \bel{secas}- \left( {\frac {d}{d\theta}}v
\left( \psi
 \right)  \right) ^{2}+{e^{2\,P_{{\infty }} \left( \psi \right) }}
 \left( {\frac {d^{2}}{d{\theta}^{2}}}Q_{{\infty }} \left( \psi
 \right)  \right) ^{2} = 0 \;,\ee the first question to answer is
how far  to push an expansion of $P$ and $Q$ to  isolate the
potentially unbounded terms in the curvature. A tedious but
straightforward inspection of the curvature tensor, as given in
the appendix, shows that terms of the form $f(\theta)t^3\ln^j|t|$
in $Q$, with $j\ne 0$, might lead to a logarithmic blow up of the
tetrad components ${R_{\mathrm{\ }(1)\,\mathrm{\ }(3)\,\mathrm{\
}(1)\,\mathrm{\ }(4 )}}$ and ${R_{\mathrm{\ }(2)\,\mathrm{\
}(3)\,\mathrm{\ }(2)\,\mathrm{\ }(4 )}}$ (here the tetrad given at
the beginning of Appendix~\ref{SA} is used), and that any higher
power of $t$ will lead to a vanishing contribution to the tetrad
components of the Riemann tensor. This means that one needs to
have the exact form of all the coefficients in an asymptotic
expansion of $Q$ up to order $O(t^3)$. Similarly one finds that
one needs to have the exact form of all the coefficients in an
asymptotic expansion of $P$ up to order $O(t^2)$. Those expansion
coefficients can be found by collecting all terms with the same
powers of $t$ and of $\ln |t|$, and setting the result to zero, in
the Gowdy equations,
\bean &\displaystyle
\partial_t^2 P -
\partial^2_\theta P = - \frac {\partial_t P }{t} + e^{2P}
\left((\partial_t Q)^2 -(\partial_\theta  Q)^2\right) \;, & \\
&\displaystyle \partial_t^2 Q - \partial^2_\theta Q = - \frac
{\partial_t Q }{t} -2 \left(\partial_t P \partial_t Q
-\partial_\theta P\partial_\theta Q\right) \;. & \eeal{Gowdyeq}
Recall that we are dealing with \avtdpqi\ solutions. It is easily
seen from \eq{avtd1}-\eq{avtd2} with $v_1(\psi)=1$ and from
\eq{Gowdyeq} that (compare~\eq{sh46a}) \bel{Qvan} (\partial_\theta
Q_\infty)(\psi)=0\;.\ee Assuming this together with \eq{secas}, we
note the following simple observations concerning the behavior of
the solution at $\theta=\psi$:
\begin{enumerate}
\item The term $Q_\infty$ does not give any undifferentiated
contribution to $P$. \item A term $f(\theta)|t|^{\alpha(\theta) }
\ln^j|t|$ in $Q$ with $\alpha$ larger than or equal to, say $7/4$,
generates in $P$, for $v(\theta)$ close to $1$, terms of the form
\bean \lefteqn{ |t|^{2v(\theta)+\alpha(\theta) -2}\Big(\hat
f(\theta)\ln^{j+1} |t| + \ \mbox{lower powers of $\ln |t|$}
\Big)}&&\\ && + \ \mbox{higher powers of $|t|$ (multiplied perhaps
by higher powers of $\ln |t|$)}\;. \nonumber \\ &&\eeal{asexp}
\item A
 term $f(\theta)|t|^{\beta(\theta)} \ln^j|t|$ in $P$ with $\beta\ge 0$
generates in $Q$, for $v(\theta)$ close to $1$, terms of the form
$\tilde f(\theta)|t|^{2v(\theta)+\beta(\theta)}\ln ^{j+1}|t|$, as
well as terms with the same power of $|t|$ but lower powers of
$\ln |t|$, or terms with higher powers of $|t|$.
\end{enumerate}
Consider the linear counterpart of \eq{Gowdyeq},
\beal{polGowdyeq} &\displaystyle \partial_t^2 f -
\partial^2_\theta f = - \frac {\partial_t f }{t}  \;. & \eea
As shown in~\cite{IMGowdy} (compare \cite[Equation~(4a)]{CIM}),
for every smooth solution of \eq{polGowdyeq} on $(-\infty,0)\times
S^1$ there exist functions
$f_{\ln}(t^2,\theta),\mathring{f}(t^2,\theta)$, which are smooth
up to boundary on the set  $(t^2,\theta)\in [0,\infty)\times S^1$,
such that
$$f(t,\theta)=f_{\ln}(t^2,\theta) \ln |t| + \mathring{f}(t^2,\theta)\;.$$
Further for any $ f_{\ln}(0,\theta)$ and $\mathring{f} (0,\theta)$
there exists a solution as above, and we have the asymptotic
expansion
\beaa f &=& f_{\ln}(0,\theta) \ln |t| +\mathring{f} (0,\theta) \\
&& +\frac{\partial_\theta^2 f_{\ln}(0,\theta)}{4} t^2 \ln
|t|+\frac{\partial_\theta^2
\mathring{f}(0,\theta)-\partial_\theta^2 f_{\ln}(0,\theta)}{4} t^2
+ O(t^4\ln |t|)\; . \eeaa Under the current hypotheses the
linearisation of the $P$ equation differs from \eq{polGowdyeq} by
terms decaying sufficiently fast so that the leading order
behavior of $P$ is correctly reflected by the above. That is not
the case anymore for $Q$, because of the $1/t$ behavior of the
$\partial_t P$ term, so that the leading terms in the $Q$ equation
linearised with respect to $Q$ are \beal{polQGowdyeq}
&\displaystyle
\partial_t^2 \tilde f -
\partial^2_\theta \tilde f = + \frac {\partial_t \tilde f }{t}  \;. & \eea
The associated indicial exponents are zero and two, from which it
is not too difficult to prove the following behavior of solutions
of \eq{polQGowdyeq}
$$ \tilde f(t,\theta)= \tilde{f}_0(t^2,\theta)+ \tilde f_{\ln}(t^2,\theta) t^2\ln |t|\;,$$
with freely prescribable functions $\tilde{f}_0 (0,\theta)$ and
$\partial_\theta^2\tilde{f}_0 (0,\theta)$, together with an
associated asymptotic expansion \beaa \tilde f &=&\tilde{f}_0
(0,\theta) +\frac{\partial_\theta^2 \tilde f_0(0,\theta)}{2} t^2
\ln |t|+ \frac{\partial_\theta^2\tilde{f}_0(0,\theta)}{2}
(0,\theta)t^2 + O(t^4\ln |t|)\; . \eeaa Since the full solution
$(P,Q)$ is already known to belong to the \avtdpqi\ class, using
what has been said one can proceed as follows: one starts with the
leading order behavior of $P$, \bel{Pstart} P= -v_1(\theta)\ln t +
P_\infty(\theta)+O_{\ln}(|t|^2)\;,\ee where $f=O_{\ln}(|t|^s)$
denotes a function which satisfies
$$|\partial_t^i\partial_\theta^jf|\le C_{i,j} |t|^{s-i} |\ln(|t|)|^{N_i,j}\;,$$
for some constants $C_{i,j}, N_{i,j}$. Inserting \eq{Pstart} into
the equation for $Q$ in \eq{Gowdyeq} one finds that at
$\theta=\psi$ we have the expansions (recall \eq{secas},
\eq{Qvan})
\bea \label{Qexp}Q(t,\theta) &=&
Q_\infty(\theta)
+\frac{t^2}2 \partial^2_\theta Q_\infty(\theta) \ln |t| +
\psi_Q(\theta) t^2 +t^4 W(t,\theta)\;.\eea Inserting \eq{Qexp} in
the first equation in \eq{Gowdyeq} one then obtains
\bean
P(t,\theta)&=&-v(\theta)\ln|t|+P_\infty(\theta)-\frac {t^2}4
\partial_\theta^2v(\theta) \ln |t|
\\ &&
+\nonumber
\frac{e^{2P_\infty(\theta)}\Big(\partial_\theta^2Q_\infty(\theta)\Big)^2}4
\Big(\ln^2 |t|-2\ln |t|+\frac 32\Big)t^2
\\ && +\nonumber
e^{2P_\infty(\theta)}\Big(4\psi_Q(\theta)+{\partial^2_\theta
Q_\infty(\theta)}\Big)\frac{\partial^2_\theta Q_\infty(\theta)}4
\Big(\ln |t|-1\Big)t^2
\\ &&
+
\left\{\frac{\partial_\theta^2P_\infty(\theta)+\partial_\theta^2v(\theta)}{4}+
e^{2P_\infty(\theta)}\Big(\psi_Q(\theta)+\frac{\partial^2_\theta
Q_\infty(\theta)}4\Big) ^2\right\} t^2
 \nonumber \\ &&
+t^4Z(t,\theta)\;.\label{Pexp}\eea All the functions appearing
above which depend only upon $\theta$ are smooth,  with $f=Z$ or
$W$ satisfying estimates of the form
$$(t\partial_t)^i (\partial_\theta)^j f = O(|\ln |t||^{j+N})$$
for some $N$. One can now insert those expansions in the  Riemann
tensor and obtain its behavior for $|t|$ small. In the case
$\frac{d v(\psi)}{d \theta}\ne 0$  but  $ (  {\frac {d}{d\theta}}v
( \psi
  )  ) ^{2}={e^{2\,P_{{\infty }} ( \psi) }}
  ( {\frac {d^{2}}{d{\theta}^{2}}}Q_{{\infty }} ( \psi
  )  ) ^{2} $, a {\sc Grtensor}
calculation gives, again at $\theta=\psi$,
\begin{equation}
C_{\alpha \beta \gamma \delta}C^{\alpha \beta \gamma \delta} \sim
3\,t \left( \ln  \left( |t| \right)  \right) ^{4}{e^{\gamma \left(
t, \theta \right) }} \left( {\frac
{d^{2}}{d{\theta}^{2}}}Q_{{\infty }}
  \left( \theta \right)  \right) ^{4} \left( {e^{P_{{\infty }} \left(
\theta \right) }} \right) ^{4}
  \label{basicccspecial}
\end{equation}
  so that with $e^{\gamma( t,\theta)} \sim 1/|t|$
then $C_{\alpha \beta \gamma \delta}C^{\alpha \beta \gamma
\delta}$ now diverges like $(\ln(|t|))^4$.

Next, if $v(\psi)=1$ and $\frac{d v(\psi)}{d \theta}=\frac
{d^{2}}{d{\theta}^{2}}Q_{{\infty }} ( \psi
  )  =0$ we
obtain
\begin{equation}
C_{\alpha \beta \gamma \delta}C^{\alpha \beta \gamma \delta} \
\sim |t| ( \ln  ( |t| )  ) ^{2}{e^{\gamma ( t,\psi
 ) }}R(\psi)
\end{equation}
as $t \rightarrow 0$, provided that $R(\psi)$  does not vanish.
Here
\begin{equation} R(\theta)= -3\, \left( {e^{P_{{\infty }} ( \theta
) }}{\frac {d^{3}}{d {\theta}^{3}}}Q_{{\infty }} ( \theta )
-{\frac {d^{2}}{d{ \theta}^{2}}}v ( \theta ) \right) \left(
{e^{P_{{\infty } } ( \theta ) }}{\frac
{d^{3}}{d{\theta}^{3}}}Q_{{\infty }}
 ( \theta ) +{\frac {d^{2}}{d{\theta}^{2}}}v ( \theta
 )  \right)\;.
\end{equation} Since  $e^{\gamma( t,\theta)} \sim 1/|t|$, we find that
$C_{\alpha
\beta \gamma \delta}C^{\alpha \beta \gamma \delta}$ diverges like
$(\ln(|t|))^2$.

One can likewise check what happens if $R(\psi)=0$:
\begin{eqnarray*} \label{limitfivespecial} C_{\alpha \beta \gamma
\delta}C^{\alpha \beta \gamma \delta} \sim -3\, ( {\frac
{d^{2}}{d{\theta}^{2}}}v ( \psi ) ) ( 2\, ( {\frac
{d}{d\theta}}P_{{\infty }} ( \psi)) ^{2}+2\,{\frac
{d^{2}}{d{\theta}^{2}}}P_{{ \infty }}( \psi) +4\,{e^{P_{{\infty }}
( \psi
  ) }}{\frac {d}{d\theta}}\psi_{{Q}}( \psi )\\\nonumber +8\,{e
^{P_{{\infty }} ( \psi) }}\psi_{{Q}}( \theta
  ) {\frac {d}{d\theta}}P_{{\infty }}( \psi ) +{
\frac {d^{2}}{d{\theta}^{2}}}v ( \psi ) ) |t|\ln
  ( |t| ) {e^{\gamma ( t,\psi) }}\;,
\end{eqnarray*}
and one obtains $\ln|t|$ behavior of the Kretschmann scalar unless
the coefficient above vanishes; in that last case the Kretschmann
scalar is bounded.

\bigskip

\appendix
\section[Appendix]
{\bf Appendix. The curvature tensor of Gowdy metrics}\label{SA}

We use the following tetrad

\begin{maplelatex}
\mapleinline{inert}{2d}{e1^a = vector([exp(1/4*gamma(t,theta))*t^(1/4), 0, 0, 0]);}{%
\[
\mathit{e1}^{a}= \left[  \! e^{(1/4\,\gamma (t, \,\theta ))}\,t^{
(1/4)}, \,0, \,0, \,0 \!  \right]
\]
}
\end{maplelatex}
\begin{maplelatex}
\mapleinline{inert}{2d}{e2^a = vector([0, exp(1/4*gamma(t,theta))*t^(1/4), 0, 0]);}{%
\[
\mathit{e2}^{a}= \left[  \! 0, \,e^{(1/4\,\gamma (t, \,\theta ))}
\,t^{(1/4)}, \,0, \,0 \!  \right]
\]
}
\end{maplelatex}
\begin{maplelatex}
\mapleinline{inert}{2d}{e3^a = vector([0, 0,
-Q(t,theta)*sqrt(exp(P(t,theta))/t),
sqrt(exp(P(t,theta))/t)]);}{%
\[
\mathit{e3}^{a}= \left[  \! 0, \,0, \, - \mathrm{Q}(t, \,\theta )
\,\sqrt{{\displaystyle \frac {e^{\mathrm{P}(t, \,\theta )}}{t}} }
, \,\sqrt{{\displaystyle \frac {e^{\mathrm{P}(t, \,\theta )}}{t} }
} \!  \right]
\]
}
\end{maplelatex}
\begin{maplelatex}
\mapleinline{inert}{2d}{e4^a = vector([0, 0, 1/(sqrt(t*exp(P(t,theta)))), 0]);}{%
\[
\mathit{e4}^{a}= \left[  \! 0, \,0, \,{\displaystyle \frac {1}{
\sqrt{t\,e^{\mathrm{P}(t, \,\theta )}}}} , \,0 \!  \right]
\]
}
\end{maplelatex}
Writing a partial derivative as a subscript, a {\sc GRTensor}
 calculation with {\sc Maple} gives the following components of the
 Riemann tensor in this frame:

\begin{maplelatex}
\mapleinline{inert}{2d}{R[``(`1`)*``(`2`)*``(`1`)*``(`2`)] =
1/4*exp(1/2*gamma)*(-exp(P)^2*t^2*Q[t]^2-2*t^3*exp(P)^2*Q[t]^2*P[t]-2*
t^3*exp(P)^2*Q[t]*Q[t,t]-t^2*P[t]^2-2*t^3*P[t]*P[t,t]-t^2*exp(P)^2*Q[t
heta]^2-2*t^3*exp(P)^2*Q[theta]^2*P[t]-P[theta]^2*t^2-1+2*t^3*P[theta,
theta]*P[t]+4*t^3*exp(P)^2*Q[t]*Q[theta]*P[theta]+2*t^3*exp(P)^2*Q[t]*
Q[theta,theta])/(t^(3/2));}{%
\maplemultiline{ {R_{\mathrm{\ }(1)\,\mathrm{\ }(2)\,\mathrm{\
}(1)\,\mathrm{\ }(2 )}}={\displaystyle \frac {1}{4}}
e^{(1/2\,\gamma )}( - (e^{P})^{2 }\,t^{2}\,{Q_{t}}^{2} -
2\,t^{3}\,(e^{P})^{2}\,{Q_{t}}^{2}\,{P_{t }} -
2\,t^{3}\,(e^{P})^{2}\,{Q_{t}}\,{Q_{t, \,t}} - t^{2}\,{P_{t}
}^{2} \\
\mbox{} - 2\,t^{3}\,{P_{t}}\,{P_{t, \,t}} - t^{2}\,(e^{P})^{2}\,{
Q_{\theta }}^{2} - 2\,t^{3}\,(e^{P})^{2}\,{Q_{\theta }}^{2}\,{P_{
t}} - {P_{\theta }}^{2}\,t^{2} - 1 + 2\,t^{3}\,{P_{\theta , \,
\theta }}\,{P_{t}} \\
\mbox{} + 4\,t^{3}\,(e^{P})^{2}\,{Q_{t}}\,{Q_{\theta }}\,{P_{
\theta }} + 2\,t^{3}\,(e^{P})^{2}\,{Q_{t}}\,{Q_{\theta , \,\theta
 }}) \left/ {\vrule height0.56em width0em depth0.56em}
 \right. \!  \! t^{(3/2)} }
}
\end{maplelatex}

\begin{maplelatex}
\mapleinline{inert}{2d}{R[``(`1`)*``(`2`)*``(`3`)*``(`4`)] =
-1/2*exp(P)*sqrt(t)*exp(1/2*gamma)*(-P[t]*Q[theta]+Q[t]*P[theta]);}{%
\[
{R_{\mathrm{\ }(1)\,\mathrm{\ }(2)\,\mathrm{\ }(3)\,\mathrm{\ }(4
)}}= - {\displaystyle \frac {1}{2}} \,e^{P}\,\sqrt{t}\,e^{(1/2\,
\gamma )}\,( - {P_{t}}\,{Q_{\theta }} + {Q_{t}}\,{P_{\theta }})
\]
}
\end{maplelatex}

\begin{maplelatex}
\mapleinline{inert}{2d}{R[``(`1`)*``(`3`)*``(`1`)*``(`3`)] =
-1/8*exp(1/2*gamma)*(-5*t*P[t]-4*t^2*P[t,t]+t^3*exp(P)^2*Q[t]^2*P[t]+t
^3*P[t]^3+t^3*exp(P)^2*Q[theta]^2*P[t]+3*t^3*P[t]*P[theta]^2+5*exp(P)^
2*t^2*Q[t]^2+t^2*P[t]^2-t^2*exp(P)^2*Q[theta]^2-P[theta]^2*t^2-1+2*t^3
*exp(P)^2*Q[t]*Q[theta]*P[theta])/(t^(3/2));}{%
\maplemultiline{ {R_{\mathrm{\ }(1)\,\mathrm{\ }(3)\,\mathrm{\
}(1)\,\mathrm{\ }(3 )}}= - {\displaystyle \frac {1}{8}}
e^{(1/2\,\gamma )}( - 5\,t\,{ P_{t}} - 4\,t^{2}\,{P_{t, \,t}} +
t^{3}\,(e^{P})^{2}\,{Q_{t}}^{2} \,{P_{t}} + t^{3}\,{P_{t}}^{3} +
t^{3}\,(e^{P})^{2}\,{Q_{\theta }
}^{2}\,{P_{t}} \\
\mbox{} + 3\,t^{3}\,{P_{t}}\,{P_{\theta }}^{2} + 5\,(e^{P})^{2}\,
t^{2}\,{Q_{t}}^{2} + t^{2}\,{P_{t}}^{2} - t^{2}\,(e^{P})^{2}\,{Q
_{\theta }}^{2} - {P_{\theta }}^{2}\,t^{2} - 1 + 2\,t^{3}\,(e^{P}
)^{2}\,{Q_{t}}\,{Q_{\theta }}\,{P_{\theta }}) \\
 \left/ {\vrule height0.56em width0em depth0.56em} \right. \!
 \! t^{(3/2)} }
}
\end{maplelatex}

\begin{maplelatex}
\mapleinline{inert}{2d}{R[``(`1`)*``(`3`)*``(`1`)*``(`4`)] =
1/8*exp(P)*exp(1/2*gamma)*(-8*t*P[t]*Q[t]+Q[t]^3*t^2*exp(P)^2+Q[t]*t^2
*P[t]^2+3*Q[t]*t^2*exp(P)^2*Q[theta]^2+Q[t]*t^2*P[theta]^2-5*Q[t]-4*t*
Q[t,t]+2*t^2*Q[theta]*P[theta]*P[t])/(sqrt(t));}{%
\maplemultiline{ {R_{\mathrm{\ }(1)\,\mathrm{\ }(3)\,\mathrm{\
}(1)\,\mathrm{\ }(4 )}}={\displaystyle \frac {1}{8}}
e^{P}\,e^{(1/2\,\gamma )}( - 8\, t\,{P_{t}}\,{Q_{t}} +
{Q_{t}}^{3}\,t^{2}\,(e^{P})^{2} + {Q_{t}}\, t^{2}\,{P_{t}}^{2} +
3\,{Q_{t}}\,t^{2}\,(e^{P})^{2}\,{Q_{\theta }
}^{2} \\
\mbox{} + {Q_{t}}\,t^{2}\,{P_{\theta }}^{2} - 5\,{Q_{t}} - 4\,t\,
{Q_{t, \,t}} + 2\,t^{2}\,{Q_{\theta }}\,{P_{\theta }}\,{P_{t}})
 \left/ {\vrule height0.41em width0em depth0.41em} \right. \!
 \! \sqrt{t} }
}
\end{maplelatex}

\begin{maplelatex}
\mapleinline{inert}{2d}{R[``(`1`)*``(`3`)*``(`2`)*``(`3`)] =
1/8*exp(1/2*gamma)*(4*t*P[t,theta]-3*t^2*P[t]^2*P[theta]-2*t^2*P[t]*ex
p(P)^2*Q[t]*Q[theta]-4*t*exp(P)^2*Q[t]*Q[theta]-P[theta]*t^2*exp(P)^2*
Q[t]^2-P[theta]*t^2*exp(P)^2*Q[theta]^2-P[theta]^3*t^2+3*P[theta])/(sq
rt(t));}{%
\maplemultiline{ {R_{\mathrm{\ }(1)\,\mathrm{\ }(3)\,\mathrm{\
}(2)\,\mathrm{\ }(3 )}}={\displaystyle \frac {1}{8}}
e^{(1/2\,\gamma )}(4\,t\,{P_{t, \,\theta }} -
3\,t^{2}\,{P_{t}}^{2}\,{P_{\theta }} - 2\,t^{2}\,{P
_{t}}\,(e^{P})^{2}\,{Q_{t}}\,{Q_{\theta }} - 4\,t\,(e^{P})^{2}\,{
Q_{t}}\,{Q_{\theta }} \\
\mbox{} - {P_{\theta }}\,t^{2}\,(e^{P})^{2}\,{Q_{t}}^{2} - {P_{
\theta }}\,t^{2}\,(e^{P})^{2}\,{Q_{\theta }}^{2} - {P_{\theta }}
^{3}\,t^{2} + 3\,{P_{\theta }}) \left/ {\vrule height0.41em
width0em depth0.41em} \right. \!  \! \sqrt{t} }
}
\end{maplelatex}

\begin{maplelatex}
\mapleinline{inert}{2d}{R[``(`1`)*``(`3`)*``(`2`)*``(`4`)] =
-1/8*exp(P)*exp(1/2*gamma)*(6*t*Q[t]*P[theta]-2*Q[t]*t^2*P[theta]*P[t]
-3*Q[t]^2*t^2*exp(P)^2*Q[theta]+4*t*Q[t,theta]-t^2*Q[theta]*P[t]^2-t^2
*Q[theta]^3*exp(P)^2-t^2*Q[theta]*P[theta]^2+3*Q[theta]+2*Q[theta]*t*P
[t])/(sqrt(t));}{%
\maplemultiline{ {R_{\mathrm{\ }(1)\,\mathrm{\ }(3)\,\mathrm{\
}(2)\,\mathrm{\ }(4 )}}= - {\displaystyle \frac {1}{8}}
e^{P}\,e^{(1/2\,\gamma )}(6\, t\,{Q_{t}}\,{P_{\theta }} -
2\,{Q_{t}}\,t^{2}\,{P_{\theta }}\,{P _{t}} -
3\,{Q_{t}}^{2}\,t^{2}\,(e^{P})^{2}\,{Q_{\theta }} + 4\,t
\,{Q_{t, \,\theta }} \\
\mbox{} - t^{2}\,{Q_{\theta }}\,{P_{t}}^{2} - t^{2}\,{Q_{\theta }
}^{3}\,(e^{P})^{2} - t^{2}\,{Q_{\theta }}\,{P_{\theta }}^{2} + 3
\,{Q_{\theta }} + 2\,{Q_{\theta }}\,t\,{P_{t}}) \left/ {\vrule
height0.41em width0em depth0.41em} \right. \!  \! \sqrt{t} }
}
\end{maplelatex}

\begin{maplelatex}
\mapleinline{inert}{2d}{R[``(`1`)*``(`4`)*``(`1`)*``(`4`)] =
1/8*exp(1/2*gamma)*(1-5*t*P[t]-4*t^2*P[t,t]+3*exp(P)^2*t^2*Q[t]^2-t^2*
P[t]^2+t^2*exp(P)^2*Q[theta]^2+P[theta]^2*t^2+t^3*exp(P)^2*Q[t]^2*P[t]
+t^3*P[t]^3+t^3*exp(P)^2*Q[theta]^2*P[t]+3*t^3*P[t]*P[theta]^2+2*t^3*e
xp(P)^2*Q[t]*Q[theta]*P[theta])/(t^(3/2));}{%
\maplemultiline{ {R_{\mathrm{\ }(1)\,\mathrm{\ }(4)\,\mathrm{\
}(1)\,\mathrm{\ }(4 )}}={\displaystyle \frac {1}{8}}
e^{(1/2\,\gamma )}(1 - 5\,t\,{P _{t}} - 4\,t^{2}\,{P_{t, \,t}} +
3\,(e^{P})^{2}\,t^{2}\,{Q_{t}}^{ 2} - t^{2}\,{P_{t}}^{2} +
t^{2}\,(e^{P})^{2}\,{Q_{\theta }}^{2}
 + {P_{\theta }}^{2}\,t^{2} \\
\mbox{} + t^{3}\,(e^{P})^{2}\,{Q_{t}}^{2}\,{P_{t}} + t^{3}\,{P_{t
}}^{3} + t^{3}\,(e^{P})^{2}\,{Q_{\theta }}^{2}\,{P_{t}} + 3\,t^{3
}\,{P_{t}}\,{P_{\theta }}^{2} + 2\,t^{3}\,(e^{P})^{2}\,{Q_{t}}\,{
Q_{\theta }}\,{P_{\theta }}) \left/ {\vrule
height0.56em width0em depth0.56em} \right. \!  \!  \\
t^{(3/2)} }
}
\end{maplelatex}

\begin{maplelatex}
\mapleinline{inert}{2d}{R[``(`1`)*``(`4`)*``(`2`)*``(`3`)] =
-1/8*exp(P)*exp(1/2*gamma)*(2*t*Q[t]*P[theta]-2*Q[t]*t^2*P[theta]*P[t]
-3*Q[t]^2*t^2*exp(P)^2*Q[theta]+4*t*Q[t,theta]-t^2*Q[theta]*P[t]^2-t^2
*Q[theta]^3*exp(P)^2-t^2*Q[theta]*P[theta]^2+3*Q[theta]+6*Q[theta]*t*P
[t])/(sqrt(t));}{%
\maplemultiline{ {R_{\mathrm{\ }(1)\,\mathrm{\ }(4)\,\mathrm{\
}(2)\,\mathrm{\ }(3 )}}= - {\displaystyle \frac {1}{8}}
e^{P}\,e^{(1/2\,\gamma )}(2\, t\,{Q_{t}}\,{P_{\theta }} -
2\,{Q_{t}}\,t^{2}\,{P_{\theta }}\,{P _{t}} -
3\,{Q_{t}}^{2}\,t^{2}\,(e^{P})^{2}\,{Q_{\theta }} + 4\,t
\,{Q_{t, \,\theta }} \\
\mbox{} - t^{2}\,{Q_{\theta }}\,{P_{t}}^{2} - t^{2}\,{Q_{\theta }
}^{3}\,(e^{P})^{2} - t^{2}\,{Q_{\theta }}\,{P_{\theta }}^{2} + 3
\,{Q_{\theta }} + 6\,{Q_{\theta }}\,t\,{P_{t}}) \left/ {\vrule
height0.41em width0em depth0.41em} \right. \!  \! \sqrt{t} }
}
\end{maplelatex}

\begin{maplelatex}
\mapleinline{inert}{2d}{R[``(`1`)*``(`4`)*``(`2`)*``(`4`)] =
-1/8*exp(1/2*gamma)*(4*t*P[t,theta]-3*t^2*P[t]^2*P[theta]-2*t^2*P[t]*e
xp(P)^2*Q[t]*Q[theta]-4*t*exp(P)^2*Q[t]*Q[theta]-P[theta]*t^2*exp(P)^2
*Q[t]^2-P[theta]*t^2*exp(P)^2*Q[theta]^2-P[theta]^3*t^2+3*P[theta])/(s
qrt(t));}{%
\maplemultiline{ {R_{\mathrm{\ }(1)\,\mathrm{\ }(4)\,\mathrm{\
}(2)\,\mathrm{\ }(4 )}}= - {\displaystyle \frac {1}{8}}
e^{(1/2\,\gamma )}(4\,t\,{P_{ t, \,\theta }} -
3\,t^{2}\,{P_{t}}^{2}\,{P_{\theta }} - 2\,t^{2}
\,{P_{t}}\,(e^{P})^{2}\,{Q_{t}}\,{Q_{\theta }} - 4\,t\,(e^{P})^{2
}\,{Q_{t}}\,{Q_{\theta }} \\
\mbox{} - {P_{\theta }}\,t^{2}\,(e^{P})^{2}\,{Q_{t}}^{2} - {P_{
\theta }}\,t^{2}\,(e^{P})^{2}\,{Q_{\theta }}^{2} - {P_{\theta }}
^{3}\,t^{2} + 3\,{P_{\theta }}) \left/ {\vrule height0.41em
width0em depth0.41em} \right. \!  \! \sqrt{t} }
}
\end{maplelatex}

\begin{maplelatex}
\mapleinline{inert}{2d}{R[``(`2`)*``(`3`)*``(`2`)*``(`3`)] =
-1/8*exp(1/2*gamma)*(t^3*exp(P)^2*Q[t]^2*P[t]+t^3*P[t]^3+t^3*exp(P)^2*
Q[theta]^2*P[t]+3*t^3*P[t]*P[theta]^2-exp(P)^2*t^2*Q[t]^2-t^2*P[t]^2+5
*t^2*exp(P)^2*Q[theta]^2+P[theta]^2*t^2-t*P[t]+1-4*t^2*P[theta,theta]+
2*t^3*exp(P)^2*Q[t]*Q[theta]*P[theta])/(t^(3/2));}{%
\maplemultiline{ {R_{\mathrm{\ }(2)\,\mathrm{\ }(3)\,\mathrm{\
}(2)\,\mathrm{\ }(3 )}}= - {\displaystyle \frac {1}{8}}
e^{(1/2\,\gamma )}(t^{3}\,(e ^{P})^{2}\,{Q_{t}}^{2}\,{P_{t}} +
t^{3}\,{P_{t}}^{3} + t^{3}\,(e ^{P})^{2}\,{Q_{\theta
}}^{2}\,{P_{t}} + 3\,t^{3}\,{P_{t}}\,{P_{
\theta }}^{2} \\
\mbox{} - (e^{P})^{2}\,t^{2}\,{Q_{t}}^{2} - t^{2}\,{P_{t}}^{2} +
5\,t^{2}\,(e^{P})^{2}\,{Q_{\theta }}^{2} + {P_{\theta }}^{2}\,t^{
2} - t\,{P_{t}} + 1 - 4\,t^{2}\,{P_{\theta , \,\theta }} \\
\mbox{} + 2\,t^{3}\,(e^{P})^{2}\,{Q_{t}}\,{Q_{\theta }}\,{P_{
\theta }}) \left/ {\vrule height0.56em width0em depth0.56em}
 \right. \!  \! t^{(3/2)} }
}
\end{maplelatex}

\begin{maplelatex}
\mapleinline{inert}{2d}{R[``(`2`)*``(`3`)*``(`2`)*``(`4`)] =
1/8*exp(P)*exp(1/2*gamma)*(Q[t]^3*t^2*exp(P)^2+Q[t]*t^2*P[t]^2+3*Q[t]*
t^2*exp(P)^2*Q[theta]^2+Q[t]*t^2*P[theta]^2-Q[t]-8*t*P[theta]*Q[theta]
+2*t^2*Q[theta]*P[theta]*P[t]-4*t*Q[theta,theta])/(sqrt(t));}{%
\maplemultiline{ {R_{\mathrm{\ }(2)\,\mathrm{\ }(3)\,\mathrm{\
}(2)\,\mathrm{\ }(4 )}}={\displaystyle \frac {1}{8}}
e^{P}\,e^{(1/2\,\gamma )}({Q_{t} }^{3}\,t^{2}\,(e^{P})^{2} +
{Q_{t}}\,t^{2}\,{P_{t}}^{2} + 3\,{Q_{
t}}\,t^{2}\,(e^{P})^{2}\,{Q_{\theta }}^{2} + {Q_{t}}\,t^{2}\,{P_{
\theta }}^{2} - {Q_{t}} \\
\mbox{} - 8\,t\,{P_{\theta }}\,{Q_{\theta }} + 2\,t^{2}\,{Q_{
\theta }}\,{P_{\theta }}\,{P_{t}} - 4\,t\,{Q_{\theta , \,\theta }
}) \left/ {\vrule height0.41em width0em depth0.41em} \right. \!
 \! \sqrt{t} }
}
\end{maplelatex}

\begin{maplelatex}
\mapleinline{inert}{2d}{R[``(`2`)*``(`4`)*``(`2`)*``(`4`)] =
1/8*exp(1/2*gamma)*(exp(P)^2*t^2*Q[t]^2+t^2*P[t]^2+3*t^2*exp(P)^2*Q[th
eta]^2-P[theta]^2*t^2+t^3*exp(P)^2*Q[t]^2*P[t]+t^3*P[t]^3+t^3*exp(P)^2
*Q[theta]^2*P[t]+3*t^3*P[t]*P[theta]^2-1-t*P[t]-4*t^2*P[theta,theta]+2
*t^3*exp(P)^2*Q[t]*Q[theta]*P[theta])/(t^(3/2));}{%
\maplemultiline{ {R_{\mathrm{\ }(2)\,\mathrm{\ }(4)\,\mathrm{\
}(2)\,\mathrm{\ }(4 )}}={\displaystyle \frac {1}{8}}
e^{(1/2\,\gamma )}((e^{P})^{2}\, t^{2}\,{Q_{t}}^{2} +
t^{2}\,{P_{t}}^{2} + 3\,t^{2}\,(e^{P})^{2}\, {Q_{\theta }}^{2} -
{P_{\theta }}^{2}\,t^{2} + t^{3}\,(e^{P})^{2}
\,{Q_{t}}^{2}\,{P_{t}} \\
\mbox{} + t^{3}\,{P_{t}}^{3} + t^{3}\,(e^{P})^{2}\,{Q_{\theta }}
^{2}\,{P_{t}} + 3\,t^{3}\,{P_{t}}\,{P_{\theta }}^{2} - 1 - t\,{P
_{t}} - 4\,t^{2}\,{P_{\theta , \,\theta }} + 2\,t^{3}\,(e^{P})^{2
}\,{Q_{t}}\,{Q_{\theta }}\,{P_{\theta }}) \\
 \left/ {\vrule height0.56em width0em depth0.56em} \right. \!
 \! t^{(3/2)} }
}
\end{maplelatex}

\begin{maplelatex}
\mapleinline{inert}{2d}{R[``(`3`)*``(`4`)*``(`3`)*``(`4`)] =
-1/4*exp(1/2*gamma)*(t^2*P[t]^2-1+exp(P)^2*t^2*Q[t]^2-P[theta]^2*t^2-t
^2*exp(P)^2*Q[theta]^2)/(t^(3/2));}{%
\[
{R_{\mathrm{\ }(3)\,\mathrm{\ }(4)\,\mathrm{\ }(3)\,\mathrm{\ }(4
)}}= - {\displaystyle \frac {1}{4}} \,{\displaystyle \frac {e^{(1
/2\,\gamma )}\,(t^{2}\,{P_{t}}^{2} - 1 + (e^{P})^{2}\,t^{2}\,{Q_{
t}}^{2} - {P_{\theta }}^{2}\,t^{2} - t^{2}\,(e^{P})^{2}\,{Q_{
\theta }}^{2})}{t^{(3/2)}}}
\]
}
\end{maplelatex}

\bigskip

\noindent{\sc Acknowledgements:} PTC wishes to thank  A.~Gerber
and J.R.~Licois for help with computer algebra calculations. He
acknowledges the hospitality of the Erwin Schr\"odinger Institute,
Vienna, during the final stage of work on this paper. Portions of
this work were made possible by use of {\sc GRTensor} \cite{grt}.


\end{document}